\documentclass[aps,twocolumn,showpacs,superscriptaddress,floatfix,letter]{revtex4}
\usepackage{graphicx}
\usepackage{psfrag}
\begin{document}
\newcommand{\beq}{\begin{equation}}
\newcommand{\eeq}{\end{equation}}
\newcommand{\ben}{\begin{eqnarray}}
\newcommand{\een}{\end{eqnarray}}
\newcommand{\bea}{\begin{array}}
\newcommand{\eea}{\end{array}}
\newcommand{\om}{(\omega )}
\newcommand{\bef}{\begin{figure}}
\newcommand{\eef}{\end{figure}}
\newcommand{\leg}[1]{\caption{\protect\rm{\protect\footnotesize{#1}}}}
\newcommand{\ew}[1]{\langle{#1}\rangle}
\newcommand{\be}[1]{\mid\!{#1}\!\mid}
\newcommand{\no}{\nonumber}
\newcommand{\etal}{{\em et~al }}
\newcommand{\geff}{g_{\mbox{\it{\scriptsize{eff}}}}}
\newcommand{\da}[1]{{#1}^\dagger}
\newcommand{\cf}{{\it cf.\/}\ }
\newcommand{\ie}{{\it i.e.\/}\ }

\title{Superradiance Transition in Transport Through Nanosystems}

\author{G.~L.~Celardo}
\affiliation{Tulane University, Department of Physics, New Orleans, Louisiana 70118}
\author{L.~Kaplan}
\affiliation{Tulane University, Department of Physics, New Orleans, Louisiana 70118}

\begin{abstract}                

Using an energy-independent non-Hermitian Hamiltonian                         
approach to open systems, we fully describe                          
transport through a sequence of potential barriers
as external barriers are varied.                                            
Analyzing the complex eigenvalues of the non-Hermitian Hamiltonian         
model, a transition to a superradiant regime is shown to occur.
Transport properties undergo a strong change at the                        
superradiance transition, where the transmission is maximized 
and a drastic change in                   
the structure of resonances is demonstrated.                               
Finally, we analyze the effect of the superradiance transition 
in the Anderson  localized regime. 
\end{abstract}                                                               
                                                                            
\date{\today}
\maketitle

\section{Introduction}

Open quantum systems are at the center of many research fields 
in physics today, ranging from quantum computing 
to transport in nanoscale and mesoscopic systems.
In particular, electronic transport in the quantum regime
can be considered one of the central subjects in
modern solid state physics~\cite{Beenakker,Lee}. 
Transport properties depend strongly on the degree of
openness of the system.
In important applications, the effect of the opening is large, 
and cannot be treated perturbatively.
Thus, a consistent way to take  
the effect of the opening into account for arbitrary coupling strength
between the system and the outside world is highly desirable.
The effective non-Hermitian Hamiltonian approach to open quantum
systems has been shown to be a very effective tool in addressing 
this issue~\cite{MW,SZNPA89,Zannals,rotter91,DittesR,Deffect2}. 

In a typical situation, we have a discrete quantum system coupled
to an external environment characterized by a continuum of states. 
Elimination of the continuum leads to an effective non-Hermitian 
Hamiltonian~\cite{MW,SZNPA89,Zannals,rotter91,DittesR}.
Analysis of the complex eigenvalues of the effective Hamiltonian
reveals a general phenomenon, namely the 
segregation of decay widths (corresponding to the imaginary part of
the complex eigenvalues).
Specifically, in a system weakly coupled to the external world,
all states tend to be similarly affected by the opening, 
but once the coupling reaches a critical value, a sharp
reconstitution of the system occurs:
almost the entire decay width is shared by a few
short-lived states, leaving all other (long-lived) 
states effectively decoupled from the external world.
The analogy between decay width segregation
and Dicke superradiance~\cite{dicke54} has been
pointed out in Refs.~\cite{SZNPA89,Zannals}, although
Dicke superradiance is associated with many-body systems, while
width segregation occurs also in the one-body case.
We will refer to this phenomenon as the 
``superradiance transition'' in the following. 
Recently, great attention has been given to translating typical
quantum optics effects, such as Dicke superradiance, into a 
solid state context~\cite{cmSR}.
In particular, the superradiance effect has been shown to occur 
in several mesoscopic systems~\cite{Deffect}.

The effective non-Hermitian Hamiltonian
approach to open systems has been used mainly under the assumptions of
Random Matrix Theory (RMT)~\cite{verbaarschot85,puebla}.
More realistic systems have also been studied, such as nuclei~\cite{Volya} 
and billiards~\cite{rotter2}. In the latter example,
segregation of resonance widths has already been demonstrated
experimentally~\cite{rotter3}. 
The effective non-Hermitian Hamiltonian technique has also been applied to
phenomenological open tight-binding models in solid state physics~\cite{Zannals,rottertb}. 
In these papers, the existence of a superradiance transition 
in such models was shown, 
but the explicit connections to realistic systems
were not considered. For instance, one might ask whether
in a realistic situation the coupling to the external environment
can be increased up to the point where a superradiance transition occurs.
Also, the energy dependence of the effective Hamiltonian
is not easy to treat exactly, so one might ask in which
realistic applications this energy dependence can be neglected.

In this paper, we consider 
the problem of transport through a sequence of potential
barriers, see Fig.~\ref{PWELL}, which
can be considered a paradigmatic model in solid state physics.
This potential profile appears in real applications, such as semiconductor
superlattices or one-dimensional arrays of quantum dots, 
and has been widely discussed 
in the literature~\cite{POT,TsuEsaki,fabry-Perot}.

The case of equally spaced potential barriers 
has been analyzed previously~\cite{POT}. 
Here a different and more general approach to the 
problem is considered. 
First, we show that for weak tunneling coupling among the wells, 
an {\it energy-independent} effective Hamiltonian approach 
produces excellent agreement 
with an exact (numerical) treatment of the problem.
Moreover, it is shown that even in this simple system a 
superradiance transition
occurs as the coupling to the external world
is increased by decreasing the widths of the external potential barriers.
With the aid of the effective Hamiltonian approach,
we recover several previous results and 
shed new light on the essential features of 
this well-studied model, allowing for
a detailed understanding of the resonance structure. 
We emphasize that the powerful effective Hamiltonian
formalism is not in any way limited to simple models of this type, 
and can be applied to situations where 
exact treatment is difficult or impossible.
In order to show this, we also analyze the case of 
random spacings among the potential barriers,
and observe the consequences of the superradiance transition in
the Anderson localization regime.

After briefly reviewing the effective Hamiltonian formalism 
in Sec.~\ref{sec_ham}, we build the effective Hamiltonian
model for a sequence of potential barriers in Sec.~\ref{sec_model}.
In Sec.~\ref{sec_sr}, the critical coupling value at which the superradiance
transition occurs is derived, and in Sec.~\ref{sec_res} we discuss
the consequences of this transition on the resonance structure.
In Sec.~\ref{sec_trans},
we show that the maximum transmission is achieved at the
superradiance transition, and we estimate analytically the exponential gain in
transmission due to the superradiance effect. 
Finally, in Sec.~\ref{sec_anderson}, 
we consider the superradiance transition in the 
Anderson localization regime, 
as a function of the disorder strength.

The effective Hamiltonian approach  
shows great promise in experimental applications,
such as quantum dots~\cite{QDOTS} and
photonic crystals~\cite{gluna}.
We also believe that the superradiance transition
can play a major role in explaining many of the 
results found in open mesoscopic  systems~\cite{QDOTS}.
even if this effect has often been neglected in the literature.

\section{Effective Hamiltonian}
\label{sec_ham}
We first sketch the essential features of the effective Hamiltonian
approach to open quantum systems. Details of the derivation can 
be found in Refs.~\cite{MW,SZNPA89,VZWNMP04, DittesR}. 

Consider a discrete quantum system described by $N$ intrinsic basis states 
$|i \rangle$ coupled to a continuum of states $|c,E \rangle$,
where $c=1\ldots M$ is a discrete quantum number labeling
$M$ channels and $E$ is a continuum quantum number representing the energy.
Let $A_i^c(E)$ be the transition amplitude between the intrinsic states
and the continuum. Then the
effective Hamiltonian for the intrinsic system, which fully takes into account
its opening to the outside, can be written as:
\begin{equation}
H_{\rm eff}(E)=H+\Delta(E)-{i \over 2} W(E)
\label{Heff1}
\end{equation}
with 
\begin{equation}
W_{ij}(E)= 2 \pi \sum_{c ({\rm open})} A_i^c(E) A_j^c(E)^* \,,
\label{W}
\end{equation}
where the sum is limited to the open channels, and
\begin{equation}
\Delta_{ij}(E)= \sum_c \, {\rm P.v.} \int dE' \, \frac{A_i^c(E') A_j^c(E')^*}{E-E'} \,.
\end{equation}
%The effective Hamiltonian takes into account the effect of the 
%opening on the intrinsic states, with
%$\Delta_{ij}(E)$ determining the energy shift while the non-Hermitian
%part $-(i/2)W_{ij}(E)$ determines the 
%decay widths.
Assuming $W_{ij}(E)$ and $\Delta_{ij}(E)$ are smooth functions of the energy,
their energy dependence can be neglected if the region of interest
is concentrated in a small energy window.
With the aid of the effective Hamiltonian, the transmission $T^{ab}(E)$ from channel
$a$ to channel $b$ can be determined:
\begin{equation}
T^{ab}(E)=|Z^{ab}(E)|^2 \,,
\label{T1}
\end{equation}
where
\begin{equation}
Z^{ab}(E)=\sum_{i,j=1}^N A_i^a \frac{1}{E-H_{\rm eff}}  (A_j^b)^*
\label{T2}
\end{equation}
is the transmission amplitude.

We can also write $T^{ab}(E)$ in a different way,
diagonalizing the effective non-Hermitian Hamiltonian $H_{\rm eff}$.
Its eigenfunctions $|r\rangle$ and $\langle \tilde{r}|$ form a
bi-orthogonal complete set,
\begin{equation}
H_{\rm eff}|r\rangle={\cal E}_{r}|r\rangle, \quad \langle \tilde{r}|
H_{\rm eff}=\langle\tilde{r}|{\cal E}^{\ast}_{r},     \label{6}
\end{equation}
and its eigenvalues are complex energies,
\begin{equation}
{\cal E}_{r}= E_{r}-\,\frac{i}{2}\,\Gamma_{r},    \label{7}
\end{equation}
corresponding to resonances centered at $E_{r}$ with widths
$\Gamma_{r}$. The decay amplitudes $A^{a}_{i}$ are transformed
according to
\begin{equation}
{\cal A}^{a}_{r}=\sum_{i}A^{a}_{i}\langle i|r\rangle, \quad
\tilde{{\cal A}}^{b}_{r}=\sum_{j}\langle \tilde{r}|j\rangle
A^{b}_{j},                                        \label{8}
\end{equation}
and the transition amplitudes are given by
\begin{equation}
Z^{ab}(E)= \sum_{r=1}^N {\cal A}^{a}_{r}\, \frac{1}{E-{\cal
E}_r}\tilde{{\cal A}}^{b}_{r} \,.                 \label{9}
\end{equation}

The complex eigenvalues ${\cal E}$ of $H_{\rm eff}$
coincide with the poles of $Z(E)$.
%The real part of the complex eigenvalues of $H_{\rm eff}$
%determines the position of the resonant peak in transmission,
%while the imaginary part determines the decay width of the 
%resonances.
It is clear that the properties of the complex eigenvalues
of the effective Hamiltonian are very important
for understanding the transport properties of the system. 

As the coupling between the intrinsic states and the external
continuum is increased, a rearrangement of the 
widths $\Gamma_r$ occurs. This rearrangement is usually referred to
as the ``superradiance'' transition.

In order to understand the origin of this transition,
we can consider a simplified version of Eq.~(\ref{Heff1}):
$H_{\rm eff}=H_0-\frac{i}{2}\gamma W$, where $\gamma$ is a parameter that 
controls the coupling strength with the external world 
(which now we assume to be of the same order of magnitude
for all the intrinsic states), and
$H_0$ is assumed to be diagonal with eigenvalues $E_0^i$.
For small $\gamma$, the first-order
complex eigenvalues of $H_{\rm eff}$ are
${\cal E}_i= E_0^i-\frac{i}{2} \gamma W_{ii}$.
%, and thus all decay widths grow linearly with $\gamma$.
If we consider the opposite limit of large $\gamma$, $H_0$ can be viewed as
a perturbation acting on $W$. Due to the factorized structure evident in
Eq.~(\ref{W}), $W$ has only $M$ non-zero eigenvalues
for $M<N$. Thus, only $M$ states will have a decay width in the
limit of large coupling, while all others will have zero width
to first order. 
Therefore, as the coupling increases, all widths initially increase
proportionally to $\gamma$, but at large couplings
only $M$ of the widths continue to increase, while the remaining
$N-M$ widths approach zero. 
This simple example suggests that a transition between these two regimes
may take place at a critical value of $\gamma$.
Roughly, the transition occurs when 
$\gamma/D \approx 1$~\cite{Zannals,rotter4,jung1,puebla},
where $D$ is the mean level spacing of $H_0$. 
Note that the qualitative criterion $\gamma/D \approx 1$ for the transition to
superradiance is valid in the case
of uniform density of states and negligible energy shift;  
when the density of states is not uniform, the transition to 
superradiance occurs as a hierarchical process~\cite{rotter4}. 
In the case of a non-negligible energy shift, see the analysis in
Sec.~\ref{sec_sr}. 

From the above discussion it should be clear that the superradiance
transition emerges in the non-Hermitian effective Hamiltonian approach
as a general phenomenon, depending not on the details of
the system, but only on the factorized structure of $W$.
 
\section{Effective Hamiltonian for a sequence of potential barriers}
\label{sec_model}

\begin{figure}[h!]
\vspace{0cm}
\includegraphics[width=7cm,angle=-90]{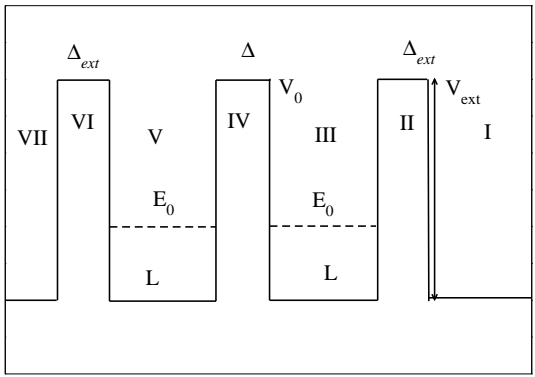}
\caption{Sequence of  potential barriers of finite height and width.
} 
\label{PWELL}
\end{figure}

Let us consider quantum transport through a sequence of $N+1$ potential 
barriers, see Fig.~\ref{PWELL}, 
of width $\Delta$, height $V_0$, and inter-barrier separation $L$.
The transport properties will be analyzed 
as we change the external barrier width $\Delta_{\rm ext}$ 
while keeping all the other barriers fixed.

We computed the transmission through this system in a standard way,
by matching the wave function and its derivative in every region,
see Fig.~\ref{PWELL}. 
Writing the wave function in region I as
$
\psi_{\rm I}=B e^{ikx}
$,
with $k=\sqrt{E}$, and in region VII as
$
\psi_{{\rm VII}}=A e^{ikx}+A' e^{-ikx}
$,
we obtain the transmission coefficient
$
T(E)=|B/A|^2
$
and the reflection coefficient
$
R(E)=|A'/A|^2=1-T(E)
$.
For comparing our numerical simulations with experimental results, 
we note that we work in $\hbar^2/2m_e=1$ units throughout. Thus, when
distances $\Delta$, $\Delta_{\rm ext}$, and $L$ in Fig.~\ref{PWELL}
are measured in nm (the typical scale in semiconductor superlattices),
all energies are calculated in units of $0.038$ eV.
In the following we set $L=2$ and $V_0=1000$.

We will now proceed to build an
effective non-Hermitian Hamiltonian to describe the quantum transport
through a sequence of potential barriers.

A sequence of $N$ potential wells
can be thought of as a closed system coupled to the continuum of
scattering states through the external barriers. 
Changing the external barrier widths or heights will change the 
coupling to the continuum.
In the limit of low tunneling coupling among the wells,
the usual tight binding approximation can be used to
model the closed system:
we define the intrinsic basis states $|i\rangle$
as the bound states in each potential well, corresponding 
to a certain energy level $E_0$. 
Each basis state is coupled to its nearest neighbor by
the tunneling coupling $\Omega$.  
For small coupling we have~\cite{landau, Gur}:
\begin{equation}
\label{Omega}
\Omega= 2 \alpha |\psi(x_0)|^2=   \frac{2 \alpha^2 E_0}{V_0 (1+\alpha L/2)} \exp{(-\alpha \Delta)} \,,
\end{equation}
where $\alpha=\sqrt{V_0-E_0}$, $k=\sqrt{E_0}$,
%$C=\sqrt{\alpha E_0/((1+\alpha L/2)V_0)}$, $A=\sqrt{\alpha/(1+\alpha L/2)}$ and
$\psi$ is a basis wave function localized in a
single potential well, and $x_0$ is a point in the middle 
of a potential barrier immediately adjacent to that well.
Due to the tunneling coupling among the $N$ wells, the 
eigenenergies of the closed system
form a miniband around $E_0$, see Eq.~(\ref{wq}) in the next Section.

The outside world is characterized by the scattering states to the
left,  $|L,E\rangle$, and to the right, $|R,E\rangle$, 
of the sequence of potential barriers.
Due to the coupling to the 
scattering states, the states $|1\rangle$ and $|N\rangle$ 
acquire a finite width $\gamma$ and an energy shift $\delta$, 
which can be computed following Refs.~\cite{Gur,QTT}.
In the case of a varying external barrier width 
$\Delta_{\rm ext}$, one obtains:

\begin{equation}
\left. \begin{array}{lll}
         \gamma = \frac{8 \alpha^3 E_0 k}{V_0^2 (1+\alpha L/2)} \exp{(-2 \alpha \Delta_{\rm ext})}\\ \\
         \delta =  \frac{k^2-\alpha^2}{4 \alpha k} \gamma\,. \\
        \end{array} \right. \label{pars}
\end{equation}
Note that the shift $\delta$ vanishes for $E_0=V_0/2$; 
otherwise the sign of $\delta$ is given by the sign of $E_0-V_0/2$.

Analogous expressions when the external potential height $V_{\rm ext}$ is varied
can be computed 
by extending the methods of~\cite{Gur}, but are more complicated
and will not be reported here. 
%Note also that we are neglecting the energy dependence of the decay widths and
%of the energy shifts. We will show later that this is a good approximation. 

We can now write the full effective Hamiltonian for the miniband
centered at energy $E_0$ as:
\begin{equation}
H_{\rm eff} =  \left( \begin{array}{ccccc}
E_0+\delta-\frac{i}{2} \gamma & \Omega & 0 & ...& 0 \\
\Omega & E_0 & \Omega & ... & 0\\
0 & \Omega & E_0 & ... & 0 \\
...&...&...&...&...\\
0& 0 & 0& ... & E_0+\delta-\frac{i}{2} \gamma \end{array} \right)
\label{Heff}
\end{equation}
Using Eqs.~(\ref{T1}) and (\ref{T2}), the transmission
through the sequence of potential barriers becomes
\begin{equation}
T(E)= \left| \frac{(\gamma/ \Omega)}{ \prod_{k=1}^N (E-{\cal E}_k)/ \Omega} \right |^2 \,.
\label{THeff}                            
\end{equation}            

From Eq.~(\ref{THeff}) we see that the spectrum of complex eigenvalues      
${\cal E}_k=E_k-\frac{i}{2} \Gamma_k$ of                                    
$H_{\rm eff}$ determines the transmission through the system.              

In order to show the range of validity of the effective Hamiltonian model,   
we compute  the normalized integrated transmission:                   
\begin{equation}
S=\frac{1}{4 \Omega}\int_{E_{\rm min}}^{E_{\rm max}} T(E) dE \,,
\label{defS}
\end{equation}
where the interval $[E_{\rm min},E_{\rm max}]$ includes the entire miniband
centered at $E_0$.

\begin{figure}[h!]
\vspace{0cm}
\includegraphics[width=7cm,angle=-90]{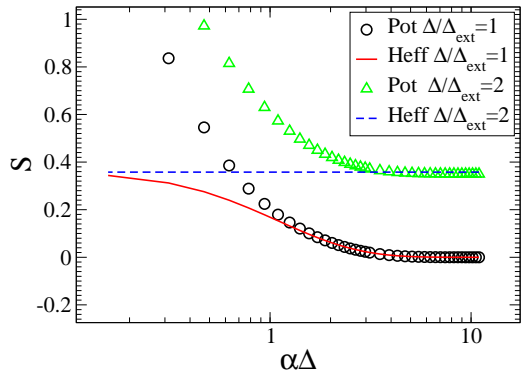}
\caption{
The integrated transmission $S$ is plotted vs $\alpha \Delta$ for 
a system with $N=10$ wells.
Different values of $\alpha \Delta$ are obtained by
varying $\Delta$ while keeping $E_0 \approx 20$ fixed.
The symbols indicate exact numerical results for
the sequence of potential barriers, while the lines represent
the effective Hamiltonian prediction. The effective Hamiltonian 
approximation is
excellent for $\alpha \Delta \gg 1$. 
Two values of $\Delta / \Delta_{\rm ext}$ are shown. 
The quantities plotted are dimensionless.
}
\label{PotvsHeffD}
\end{figure}

The predictions of  Eq.~(\ref{THeff}) are now compared with
the exact numerical results.
The effective Hamiltonian approach is expected to break down for small values
of $\alpha \Delta$. 
In Fig.~\ref{PotvsHeffD}, we plot $S$ vs $\alpha \Delta$
for a system of $N=10$ wells, with $E_0 \approx 20$ fixed, 
so that $\alpha$ remains constant. From the figure we can see that the
effective Hamiltonian approximation gives excellent results for $\alpha \Delta \gg 1$. 
Note also that $S$ is independent
of $\Delta$ for $\Delta/\Delta_{\rm ext}=2$ in the weak coupling limit, $\alpha \Delta \gg 1$,  
as indicated by a dashed line in Fig.~\ref{PotvsHeffD}. 
This follows from the fact that $\gamma/ \Omega$ 
is independent of $\Delta$ when $\Delta/ \Delta_{\rm ext}=2$, 
see Eq.~(\ref{Omega}) and Eq.~(\ref{pars}).
Note that we also compared the results obtained with the effective
Hamiltonian approach with available analytical results
found in the literature \cite{POT}, and found excellent agreement
in the regime  $\alpha \Delta \gg 1$. 

\section{Superradiance Transition}
\label{sec_sr}

We will now analyze the superradiance transition that
occurs in the effective Hamiltonian model built in the previous section.

Diagonalization of the intrinsic Hamiltonian leads to
the energy levels~\cite{Zannals}:
\begin{equation}
w_q=E_0-2 \Omega \cos(\pi q/(N+1)) \,,
\label{wq}
\end{equation}
with $q=1 \ldots N$. 
Due to coupling with the external world, the energy levels
$w_q$ will acquire decay widths $\Gamma_q$. These decay widths are
the imaginary parts of the eigenvalues
of the effective Hamiltonian, Eq.~(\ref{Heff}),
and, for $\gamma \ll 1$, they can be written as:
\begin{equation}
\Gamma_q= \frac{4 \gamma}{N+1}  \sin^2(\pi q/(N+1)) \,.
\label{gg}
\end{equation}
We see that all widths increase proportionally to $\gamma$ for small coupling.
In the opposite limit of large $\gamma$, only $M$ states (where $M$ is the number of channels)
will have a width proportional to $\gamma$, while the 
widths of the remaining states fall off as $1/ \gamma$, as explained above in Sec.~\ref{sec_ham}. 
In our case we have $M=2$, corresponding to one scattering channel each on the
left and right.
%These two superradiant states are formed as
%linear combinations of the intrinsic basis states $|1\rangle$ and
%$|N\rangle$. This can be shown considering the matrix 
%$W$, see Eq.~(\ref{Heff1}), in the effective Hamiltonian.
The two superradiant states correspond to the
two non-zero eigenvalues of the matrix $W$ (Eq.~(\ref{Heff1})).

\begin{figure}[h!]
\vspace{0cm}
\includegraphics[width=7cm,angle=-90]{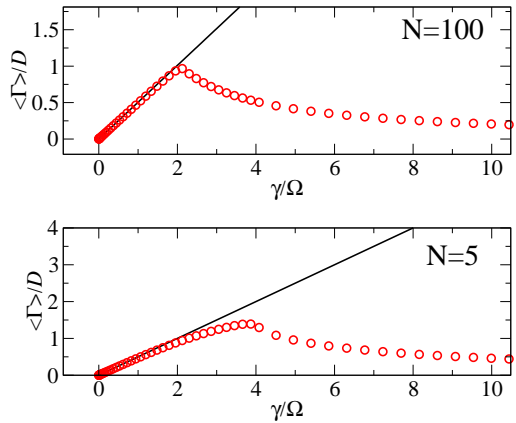}
\caption{
The average width, normalized to the mean level spacing, 
is shown as a function of 
$\gamma/\Omega$ for the case $\Delta=0.15$
and $E_0  \approx V_0/2$. 
When $N$ is large, the transition to superradiance 
is shown to occur at $\gamma/\Omega \approx 2$,
in agreement with the analytical estimation, see text.
The solid line corresponds to an average over all $N$
widths, while the symbols are obtained by averaging over
the $N-2$ smallest widths. The quantities plotted are dimensionless.
} 
\label{GD}
\end{figure}

In order to find the critical value of the parameter $\gamma$ at which the 
superradiance transition occurs,
we may analyze the average width $\langle \Gamma \rangle$ 
of the $N-M$ narrowest widths as a function 
of the coupling $\gamma$, Fig.~\ref{GD}.
At the critical value of $\gamma$, 
the average width $\langle \Gamma \rangle$ peaks and begins to decrease.
This is the signature of the superradiance transition.

We can evaluate this critical value of $\gamma$ using the criterion 
discussed earlier in Sec.~\ref{sec_ham}, $ \langle \Gamma \rangle /D \approx 1$.
Consider first the simpler case of vanishing energy shift $\delta$.
The average width $\langle \Gamma \rangle$ is then given by the perturbative expression,
Eq.~(\ref{gg}), taking into account that
$\langle \sin^2(\pi q/ (N+1))\rangle \to 1/2$ for large $N$. Moreover,
from Eq.~(\ref{wq}) we find that for large $N$ the mean level spacing becomes
$D=4 \Omega/N$, so we obtain: 
\begin{equation}
\frac{\langle \Gamma \rangle}{D}= \frac{1}{2} \frac{ \gamma}{\Omega} \,.
\end{equation}
Thus, for $\delta=0$, the criticality criterion  $ \langle \Gamma \rangle/D \approx 1$ 
implies $\gamma \approx 2 \Omega$. Note that this happens when
$\Delta_{\rm ext}=\Delta/2$ (see Eqs.~(\ref{Omega}) and (\ref{pars})),
so that the superradiance transition occurs when the external barriers
are precisely half as wide as the internal ones.

Typical examples for large and small $N$ are presented in Fig.~\ref{GD},
where $E_0=V_0/2$ to ensure that the energy shift $\delta=0$.
The $N=100$ example in the upper panel illustrates
that the estimate $\gamma \approx 2 \Omega$ for the
critical value works very well at large $N$.

We now turn to the $\delta \ne 0$ case.
Both the density of states 
and the resonance widths are modified, as we can see using
second order perturbation theory in small $\gamma$:
\begin{equation}
{\cal E}_q=w_q+(\delta -i \gamma/2)\frac{4 \sin^2{\phi_q}}{N+1} 
+(\delta^2-\gamma^2/4 -i \gamma \delta)\sum_{p \ne q} \frac{A_{qp}^2}{w_q-w_p} \,,
\label{2nd}
\end{equation}
where 
$
A_{qp}=(2/N)^2 (1+(-1)^{q+p})^2 \sin^2{\phi_q} \sin^2{\phi_p}
$
and $\phi_q=\pi q/(N+1)$.

Clearly, the local level spacings
$\Delta_q(\gamma)={\rm Re}({\cal E}_q-{\cal E}_{q-1})$
and the local resonance widths 
$\Gamma_q(\gamma)=-{\rm Im}({\cal E}_q+{\cal E}_{q-1})$
depend on the index $q$ as well as the coupling $\gamma$. 
A reasonable hypothesis is that
the superradiance transition 
occurs when the resonances begin to overlap locally, i.e.,
$\Gamma_q(\gamma) \approx D_q(\gamma)$ for some $q$.
We have confirmed numerically that this local overlap criterion
gives an excellent approximation for the critical value of
$\gamma$ at which the superradiance transition occurs, for any $\delta$. 
Unfortunately, second order 
perturbation theory does not provide an accurate analytical estimate for
$\gamma$, confirming that the physics 
is highly non-perturbative near the superradiance transition.

\section{Resonance Structure}
\label{sec_res}

To show the consequences of the
superradiance transition on the transport properties, here we 
analyze the resonance structure, by considering
the transmission $T(E)$. 
Note that the resonance structure can be directly resolved experimentally, 
see~\cite{fabry-Perot}.

\begin{figure}[h!]
\vspace{0cm}
\includegraphics[width=7cm,angle=-90]{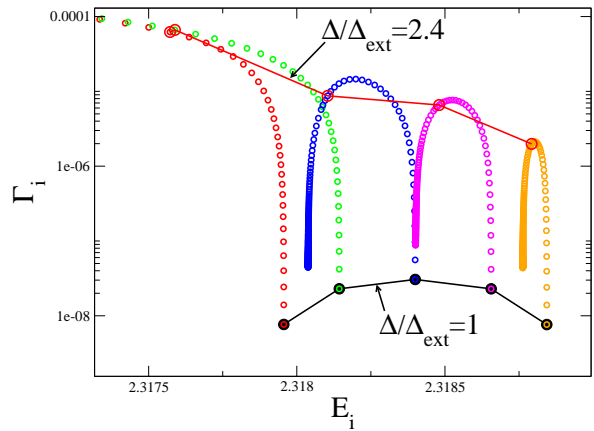}
\caption{
The evolution in the complex plane of the eigenvalues of
the effective Hamiltonian is shown as the ratio $\Delta/ \Delta_{\rm ext}$
is varied. A system of $5$ intrinsic states in the $E_0 \approx 2$ miniband is considered, with $\Delta=0.2$.
The emergence of two superradiant states
is clearly visible above the transition. Here and in the following figures, we use units $\hbar^2/2m_e=1$, so that
if lengths $\Delta$, $\Delta_{\rm ext}$, and $L$ are measured in nm, energies are calculated in units
of $0.038$ eV.}
\label{5complex}
\end{figure}

In this section, we focus on the case of $N=5$ potential wells.
As discussed above in Sec.~\ref{sec_sr}, a signature of the superradiance transition
is the segregation of resonance widths above 
the critical coupling. The system
under study has two open channels, thus we expect
two resonance widths (associated with superradiant states)
to continue increasing above the transition,
while the remaining widths approach zero.
In Fig.~\ref{5complex}, we show the trajectories of 
the complex eigenvalues ${\cal E}_i$ of the effective Hamiltonian 
as $\Delta/ \Delta_{\rm ext}$ is increased. 
Note that the real parts of the 
eigenvalues experience a leftward shift with increasing coupling,
since in this case
$E_0< V_0/2$, so that $\delta<0$.

\begin{figure}[h!]
\vspace{0cm}
\includegraphics[width=7cm,angle=-90]{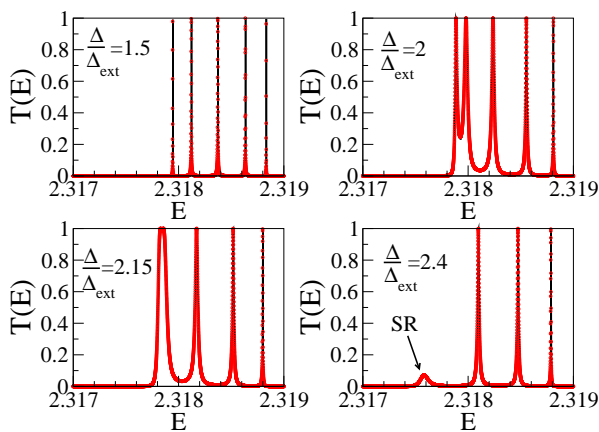}
\caption{
The transmission is shown as a function of energy for a system of
$N=5$ potential wells, for the same parameters as in
Fig.~\ref{5complex}, and several values of $\Delta/\Delta_{\rm ext}$.
The exact transmission, black solid line, is compared with the 
result obtained from the effective Hamiltonian model,
Eq.~(\ref{THeff}), indicated by red circles.
The two superradiant states merge at $\Delta/\Delta_{\rm ext}=2.15$,
and then disappear entirely, leaving behind $N-2$ resonances, 
see $\Delta/\Delta_{\rm ext}=2.4$.
} 
\label{N5}
\end{figure}

In Fig.~\ref{N5}, the transmission is shown as a function of the
energy for several values of $\Delta/ \Delta_{\rm ext}$. 
For small coupling, $\Delta/ \Delta_{\rm ext}=1.5$, we have 
$N=5$ narrow resonances as expected.
As we decrease the external barrier widths, the transmission
increases and  near $\Delta/ \Delta_{\rm ext}=2$ 
two of the resonances start to overlap. At 
$\Delta/ \Delta_{\rm ext}=2.15$, they merge, forming a broad resonance.
This shows that the superradiance transition
has a clear signature in
the resonance structure.
As the external barrier widths continue to decrease, 
the height of the superradiant resonance decreases (see
the case $\Delta/ \Delta_{\rm ext}=2.4$ in Fig.~\ref{N5}), 
until it disappears entirely for large $\Delta/ \Delta_{\rm ext}$,
due to destructive interference between the two superradiant states.
In this limit, $N-2$ narrow resonances remains. 
The fact that two resonances disappear for very large coupling is
not surprising; indeed in the absence of the two external barriers,
we simply have a system of $N-2$ wells. What is interesting is that
the two individual resonances disappear 
long before the external barriers vanish,
indeed immediately after the superradiance transition. 
Note also in Fig.~\ref{N5} that
results obtained from the effective Hamiltonian
model are indistinguishable from numerical results obtained 
by matching the wave functions.

The behavior we have demonstrated for the case of $N=5$ 
potential wells generalizes easily to a larger number of wells,
with important quantitative differences.
Indeed, for a longer chain, the superradiant states 
disappear much faster as we increase $\Delta/ \Delta_{\rm ext}$
above the critical value, i.e., the superradiance transition becomes increasingly sharp.
Interestingly, the critical value of $\Delta/ \Delta_{\rm ext}$ 
becomes both $N$-independent and $E_0$-independent in the large-$N$ limit,
with the transition occurring at $\Delta/ \Delta_{\rm ext}=2$,
in agreement with previous results~\cite{POT,fabry-Perot}.
%Note that this critical value was derived in the previous section only for the
%case $E_0=V_0/2$ for which $\delta=0$. 

\section{Integrated Transmission}
\label{sec_trans}

\begin{figure}[h!]
\vspace{0cm}
\includegraphics[width=7cm,angle=-90]{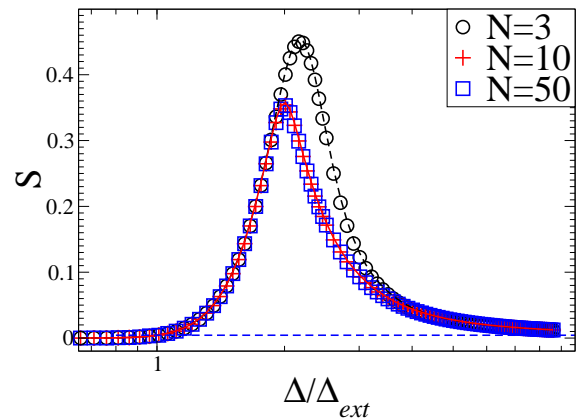}
\caption{
The integrated transmission $S$ is shown as a function of
 $\Delta / \Delta_{\rm ext}$, 
for different numbers of wells. 
The symbols refer to an exact numerical calculation, 
while the effective Hamiltonian result is indicated by
the black and red dashed curves. 
The horizontal dashed line represents the 
value of $S$ for $\Delta=\Delta_{\rm ext}$. 
In this example, we use $E_0 \approx 20$ and $\Delta=0.15$.
} 
\label{Super}
\end{figure}

Another interesting quantity to analyze is the 
integrated transmission $S$, Eq.~(\ref{defS}),
as a function of $\Delta / \Delta_{\rm ext}$.
The quantitative enhancement in $S$  when external barrier parameters are adjusted
is important in applications, for instance in the
design of electron band-pass filters for semiconductor superlattices~\cite{POT}.
From Fig.~\ref{Super}, we see that 
$S$ reaches a maximum as a function of the external barrier width.
Fig.~\ref{Super} also shows that  the value of $S$ 
for $\Delta_{\rm ext} \ll \Delta$
is the same as for $\Delta_{\rm ext}=\Delta$.
This is due to the fact that $S$ becomes $N$-independent
for large $N$, as shown in Fig.~\ref{Super}:
as the external barriers disappear 
we are eventually left with a sequence of $N-2$ potential wells, 
which has the same value of the integrated transmission $S$ as the original
sequence of $N$ wells.

\begin{figure}[h!]
\vspace{0cm}
\includegraphics[width=7cm,angle=-90]{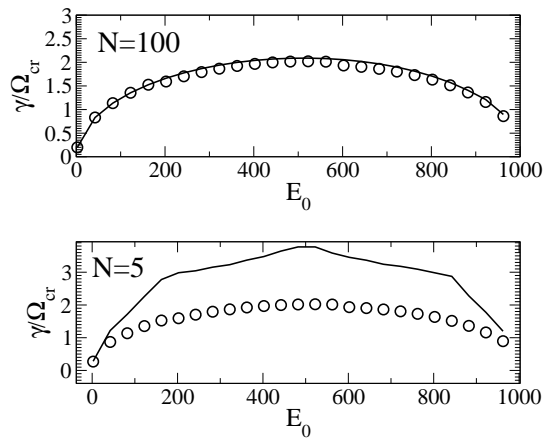}
\caption{
The critical value of $\gamma/ \Omega$ at which the superradiance transition
occurs (solid curve) is compared with the critical value at which 
the integrated transmission has a maximum (circles).
For $N=100$ (upper panel), the two coincide, while they
differ for $N=5$ (lower panel), see the discussion in the text.
Here $\Delta=0.2$.
}
\label{GE0}
\end{figure}

\begin{figure}[h!]
\vspace{0cm}
\includegraphics[width=7cm,angle=-90]{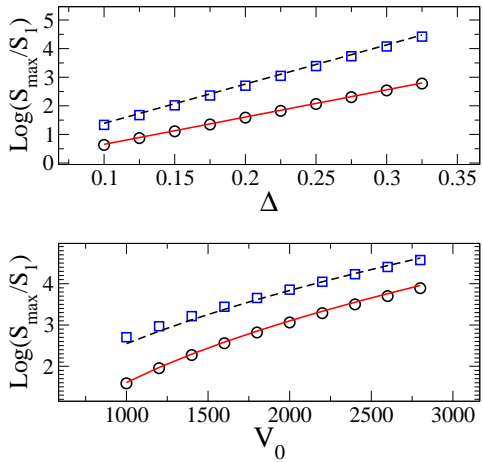}
\caption{
The transmission gain $S_{\rm max}/S_1$ is shown as a function 
of the internal barrier width $\Delta$
(upper panel) and as a function of the internal barrier height 
$V_0$ (lower panel). 
In each panel, numerical results obtained for 
energy $E_0\approx 500$ (circles) and $E_0\approx 2$ (squares)
are compared with the
analytical formula (\ref{Smax}), indicated by the curves. In the upper panel,
we fix $V_0=1000$, while in the lower panel we fix $\Delta=0.2$. 
All data points are for a system of $N=10$ wells.
The numerical prefactor in the analytical result, Eq.~(\ref{Smax}), is
chosen to fit the numerical data.
} 
\label{SmaxS0}
\end{figure}

The maximum of the integrated transmission can be related to the
superradiance transition. 
In Fig.~\ref{GE0}, we show the critical value of $\gamma / \Omega$ 
at which the average width $\langle \Gamma \rangle$ has a maximum 
(signaling the superradiance transition, see Fig.~\ref{GD}), 
compared with the  value of $\gamma / \Omega$ at which the transmission $S$
has a maximum. For large $N$, the transmission maximum is reached
precisely at the superradiance transition, while for small $N$
the specific structure of the resonances influences 
the exact position of the maximum. 
The relationship between the superradiance transition and 
the transmission maximum
can be explained as follows: for small $\gamma$, 
the resonance widths increase with $\gamma$, and so does the integrated transmission $S$.
Once the superradiant states start to form, they interfere
destructively and the associated resonances
disappear, while the widths of the other
resonances decrease with $\gamma$, leading to an overall falloff in $S$.

In order to estimate the transmission ``gain,'' we compute the ratio of the
maximum transmission to the transmission at
$\Delta/ \Delta_{\rm ext}=1$: $S_{\rm max}/S_{1}$. The transmission is 
given by the area under the resonances, and for isolated resonances
we can estimate $S \approx 2N  \langle\Gamma\rangle/ 4 \Omega$. 
Since for small $\gamma$, 
see Eq.~(\ref{gg}), we have $\langle\Gamma\rangle =2\gamma/(N+1)$, 
we can write: 
\begin{equation}
S=\frac{\gamma}{\Omega} \frac{N}{N+1} \approx \frac{ \gamma}{\Omega},
\label{S}
\end{equation}
which is $N$-independent for large $N$ as noted above. 
Taking into account that the superradiance transition 
occurs at $\gamma=2 \Omega$ for $E_0=V_0/2$, the gain can be estimated as:
\begin{equation}
 \frac{S_{\rm max}}{S_{1}}\propto \frac{\Omega}{\gamma_{1}} = \frac{V_0}{4 \alpha \sqrt{E_0}} e^{\alpha \Delta} \,.
\label{Smax}
\end{equation}
In Fig.~\ref{SmaxS0}, we show that our estimate works very well.
Eq.~(\ref{Smax}) is in agreement with the results obtained in~\cite{POT}
using a different approach. 
Note also that the gain is exponential in the internal barrier width
$\Delta$.

\section{Anderson Localization Regime}
\label{sec_anderson}

In the previous sections we considered 
an effective Hamiltonian (\ref{Heff}) with equal diagonal energies
$E_0$. Here we want to apply the effective Hamiltonian technique to
the case of random variations of the
diagonal energies: $E_0 \pm \delta E_0$, 
where $\delta E_0$ is a random variable uniformly distributed
in $[-W/2 ,+W/2]$, and $W$ is a disorder parameter.
A first analysis of this model can be found in Ref.~\cite{varenna06}.

Random variation in the diagonal energies can be thought of as a 
consequence of small random fluctuations
$\delta L$ of the well widths $L$.
For $E_0 \ll V_0$, the eigenenergies 
of a finite potential well may be approximated by the eigenenergies of 
an infinite potential well, $E_0= n^2 \pi^2/ L^2$, 
where $n=1,2,...$. For small fluctuations
$\delta L /L \ll 1$, we have
\begin{equation}
\delta E_0 =    \frac{2n^2 \pi^2}{L^3} \delta L= -C \delta L \,,
\label{en}
\end{equation}
where $C=2  n^2 \pi^2 /L^3$.  
Thus, a random variation of $\delta E_0$ in $[-W/2,+W/2]$ corresponds
to a random variation of $\delta L$ in $[-W/2C, +W/2C]$.

The effective non-Hermitian Hamiltonian with diagonal disorder 
is equivalent to an open Anderson tight binding 
model~\cite{Anderson, Lee}.
The eigenstates of the Anderson model
are exponentially localized on the system sites, with exponential
tails given by $\exp(-x/L_{\rm loc})$, where for weak disorder,
the localization length $L_{\rm loc}$ at the center of the energy band
can be written as~\cite{Felix}:

\begin{equation}
L_{\rm loc} \approx 105.2 \left(\frac{W}{\Omega}\right)^{-2} \,.
\label{loc}
\end{equation}

For $L_{\rm loc}\ll N$, the transmission decays exponentially with $N$;
this is the localized regime.
Note that for zero disorder, the transmission is $N$-independent,
as we showed in the previous section.
The condition $L_{\rm loc}=N$ defines a critical value of $(W/ \Omega)_{\rm cr}$
for the localized regime, at any given $N$.  
In the localized regime, the transmission is log-normally distributed,
and we have~\cite{Beenakker}:

%{\bf *** what does proportionality constant depend on? And why do you include 2 if it's proportional?}
\begin{equation}
\langle -\ln T\rangle = 2 \frac{N}{L_{\rm loc}} +{\rm Const}\,. 
\label{Tloc}
\end{equation}

\begin{figure}[h!]
\vspace{0cm}
\includegraphics[width=7cm,angle=-90]{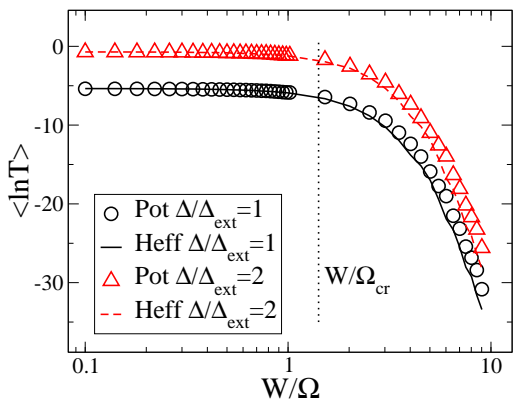}
\caption{
%{\bf *** check minus sign in y axis label? Also check same in the next figure}
The average of the logarithm of the transmission is plotted
as a function of $W/\Omega$ for $\Delta/ \Delta_{\rm ext}=1$
and $\Delta/ \Delta_{\rm ext}=2$. 
Results obtained from the effective Hamiltonian (curves)
are compared with results obtained numerically for the sequence
of potential barriers (symbols). Here we take $N=50$,
$E_0 \approx 20$, and $\Delta=0.2$. The dotted vertical line indicates
the critical value of
$W/ \Omega$, obtained from the condition
$L_{\rm loc}=N$, where $L_{\rm loc}$ is given by Eq.~(\ref{loc}).
} 
\label{PotvsHeffW}
\end{figure}

In Fig.~\ref{PotvsHeffW}, we show the average transmission 
as a function of the disorder strength for two different degrees of opening
of the system.  
The results obtained using an
effective Hamiltonian with diagonal disorder
are compared with numerical simulations for 
the disordered  sequence of potential wells.
The agreement is excellent up to a large value of the disorder,
where of course our approximations break down. [Indeed 
$\Omega$, Eq.~(\ref{Omega}), has been computed assuming that the 
energy levels are aligned, which is not true anymore 
in the presence of disorder.]
We stress that Eq.~(\ref{THeff}) for the transmission
remains valid even in the disordered case, making the use of the 
effective Hamiltonian formalism very efficient, since 
only the eigenvalues
of the effective Hamiltonian are needed to obtain the transmission.

\begin{figure}[h!]
\vspace{0cm}
\includegraphics[width=7cm,angle=-90]{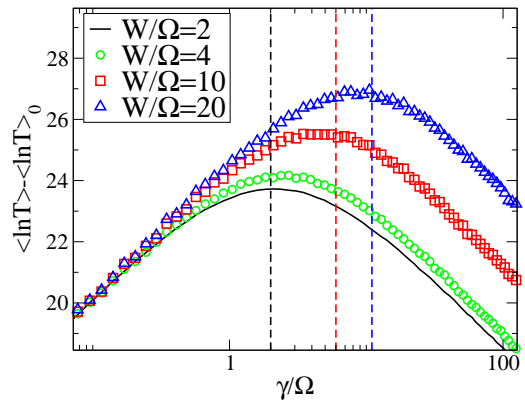}
\caption{
The mean logarithm of the transmission is plotted
as a function of $\gamma/ \Omega$,
for different values of the disorder parameter $W/\Omega$. 
Here we take $N=100$, $E_0 \approx V_0/2$, and $\Delta=0.6$.
The dashed vertical lines indicate the predicted
value of  $\gamma/ \Omega$
at which the transmission maximum is expected for $W/\Omega=2$, $10$, and $20$,
see the discussion in the text. Note that $\langle\ln T\rangle_0$ stands for
the value of $\langle \ln T\rangle$ at $\Delta=\Delta_{\rm ext}$.
} 
\label{nr10^5}
\end{figure}

The phenomenon of Anderson localization
was studied in a closed disordered chain or for fixed opening, while
in our case we can vary the degree of openness of the system.
The effect of the opening on Anderson localization is not obvious.
Will a maximum of the transmission still exist as we vary the coupling
of the system with the external world? 
Will the localization length change as we 
open up the system? 
To answer these questions, we have analyzed the effective Hamiltonian,
neglecting the role of the energy shift, i.e., we set $\delta=0$
in the following.
In Fig.~\ref{nr10^5}, we compute the average of $-\langle \ln T\rangle$ over
$10^5$ realizations as a function of $\gamma/ \Omega$. 
The energy is fixed at $E=E_0$.
%{\bf *** what is $E=0$?}
As  we vary $\gamma/ \Omega$,
the average transmission reaches a maximum, just as in the disorder-free case.

Interestingly, as the disorder strength increases,
the transmission maximum (associated with the
superradiance transition) shifts to ever higher values of the coupling strength $\gamma$.
Indeed, the mean level spacing $D$ increases with growing disorder, so that
the condition $\langle \Gamma \rangle/D \approx 1$ for the superradiance transition to occur
 will be satisfied at increasingly
larger values of $\gamma/ \Omega$.
For weak disorder, the disorder-induced correction to $D$ is
second order in the disorder strength, so $(\gamma/\Omega)_{\rm cr}=2+O(W^2/\Omega^2)$.
In the opposite regime of large $W/\Omega$ we find
 $D \approx 2\Omega(1+W/2\Omega)/N$, and the critical coupling is predicted to
 be $(\gamma/\Omega)_{\rm cr} \approx 1+W/2\Omega$.
This estimate works quite well, as shown by the vertical lines in Fig.~\ref{nr10^5}.
Note that the curves shown in  Fig.~\ref{nr10^5}
have been found to be independent of $N$. 

\begin{figure}[h!]
\vspace{0cm}
\includegraphics[width=7cm,angle=-90]{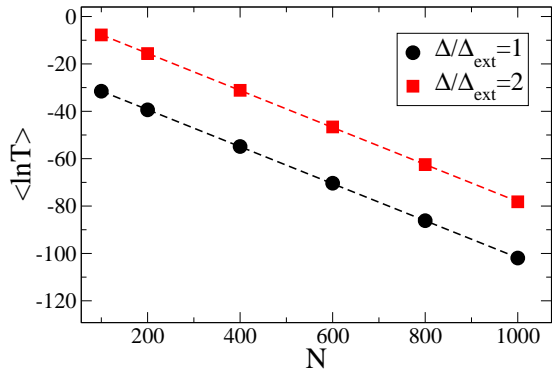}
\caption{
The mean logarithm of the transmission is plotted
as a function of $N$ for $\Delta/ \Delta_{\rm ext}=1$ (black circles) and $\Delta/ \Delta_{\rm ext}=2$
(red squares) for the case $W/\Omega=2$, $E_0 \approx V_0/2$ and $\Delta=0.6$. 
The remaining parameters are the same as in Fig.~\ref{PotvsHeffW}.
From the fit we have $\langle \ln T\rangle= a-\nu N$, with $\nu \approx 0.078$,
for both $\Delta/ \Delta_{\rm ext}=1$, $2$, 
in good agreement with the theoretical result
$\nu=2/L_{\rm loc}=0.076$, see Eq.~(\ref{Tloc}).
} 
\label{1Dloc}
\end{figure}

Finally in  Fig.~\ref{1Dloc}, we show $\langle \ln  T\rangle $ versus $N$ for $W/ \Omega=2$,
and for two different values of the external barrier width:
$\Delta/ \Delta_{\rm ext}=1$ and $\Delta/ \Delta_{\rm ext}=2$, 
where the maximum of the transmission occurs.
Fig.~\ref{1Dloc} shows that Eq.~(\ref{Tloc}) works very well in both situations, even though
the transmission is enhanced when $\Delta/ \Delta_{\rm ext}=2$,
for all values of $N$.
Thus, the localization length in a disordered 1D model is not 
affected by the opening, but the transmission is.
%From the results shown in Fig.~\ref{nr10^5} and Fig.~\ref{1Dloc},
%we can express the average transmission in the localized regime as
%$\langle \ln T\rangle=F(\gamma/\Omega) -2N/L_{\rm loc}$.

\section{Conclusion}
                           
We have analyzed quantum transport through a finite sequence of potential barriers,
a paradigmatic model for transport in solid state physics.
In this paper, the effective non-Hermitian
Hamiltonian approach has been used to analyze
the transmission through this class of systems.
The main results of our work are the following:
$i)$ we show that for weak or moderate tunneling coupling
among the potential wells, the system is well described by
an energy-independent effective Hamiltonian. 
Knowledge of the complex eigenvalues of the effective Hamiltonian
is sufficient to study transport through the system.
$ii)$ As the coupling to the
continuum is increased by adjusting the width of the external barriers,     
a superradiance transition (or Dicke effect) occurs.   
Analysis of the complex eigenvalues of the effective Hamiltonian
allows us to determine the critical coupling associated with this
transition.
$iii)$ The superradiance transition 
has strong effects on the transport properties: 
specifically, the transmission through the system is maximized at 
the superradiance transition. An expression for the 
 transmission gain due to the superradiance transition
is derived.
Moreover the resonance structure is drastically affected: at 
the superradiance transition, we have  the formation of a
broad resonance corresponding to the  superradiant states.
Beyond the transition, this broad resonance
disappears, and the number of resonances 
decreases by two.
$iv)$ The case of a disordered sequence of potential barriers 
has been also analyzed. In the presence of disorder,
Anderson localization occurs.
We have shown the localization length remains constant as the opening changes,
but the transmission has a maximum 
as a function of the coupling to the external world.
The critical value of the coupling
increases with the degree of disorder, and we obtain an estimate of the critical value
for strong and weak disorder,
based on the superradiance mechanism.

In the future it will be interesting to study the consequences of
the superradiance transition beyond the single particle
approximation, where electron-electron interactions play an important role.

\section*{Acknowledgments}

We acknowledge useful discussions with G.~P.~Berman, F.~Borgonovi,
F.~Izrailev, S.~Sorathia, and V.~G.~Zelevinsky. This work was supported
in part by the NSF under Grant No. PHY-0545390.

%%%%%%%%%%%%%%%%%%%%

\end{document}